\documentclass[12pt]{iopart}
\usepackage{iopams}

\usepackage{graphicx}
\usepackage{url}

\newcommand{\be}{\begin{equation}}
\newcommand{\ee}{\end{equation}}
\newcommand{\ba}{\begin{array}}
\newcommand{\ea}{\end{array}}
\newcommand{\bea}{\begin{eqnarray}}
\newcommand{\eea}{\end{eqnarray}}
\newcommand{\ket}[1]{| #1 \rangle}

\newcommand{\supp}{{\rm Supp}}

\newcommand{\calC}{{\cal C }}

\newcommand{\calP}{{\cal P }}

\newcommand{\calS}{{\cal S }}

\newcommand{\calF}{{\cal F }}

\newcommand{\la}{\langle}
\newcommand{\ra}{\rangle}

\newcommand{\nn}{\nonumber}

\newcommand{\mjr}[1]{\mathrm{Maj}{(#1)}}
\newcommand{\smaj}{{\calS}_{\mathrm{maj}}}
\newcommand{\Call}{C_{\rm all}}

\newtheorem{lemma}{Lemma}
\newtheorem{prop}{Proposition}

\newtheorem{corol}{Corollary}

\begin{document}

\title{Majorana Fermion Codes}

\author{Sergey Bravyi and Barbara M. Terhal}
\address{IBM Watson Research Center, Yorktown Heights, NY 10598, USA}
\author{Bernhard Leemhuis}
\address{Institute for Theoretical Physics, University of Amsterdam, Valckenierstraat 65,
1018 XE Amsterdam, The Netherlands}
\begin{abstract}

We initiate the study of Majorana fermion codes. These codes can be viewed as extensions of Kitaev's 1D model of unpaired Majorana fermions in quantum wires to higher spatial dimensions and interacting fermions.
 The purpose of Majorana fermion codes (MFCs) is to protect quantum information against low-weight fermionic errors,
 that is, operators acting on sufficiently small subsets of fermionic modes. We examine to what extent MFCs can surpass
  qubit stabilizer codes in terms of their stability properties. A general construction of 2D MFCs is proposed which combines topological protection based on a macroscopic code distance with  protection based on fermionic parity conservation. Finally, we use MFCs to show how to transform any qubit stabilizer code
to a weakly self-dual CSS code.

\end{abstract}

\maketitle

\section{Introduction}

The physical realization of systems with topological quantum order and their theoretical description
have been a topic of significant attention lately.
This attention is partly motivated by the potential use of
topologically ordered systems as fault-tolerant hardware in a quantum computer~\cite{kitaev:anyons}.
Encoding of quantum information into the ground states of such systems permits exponential suppression of dephasing,
while the presence of a gap above the ground state suppresses thermal error excitations.
The original insights~\cite{kitaev:anyons,Wen90} concerning the zero-temperature stability of topologically ordered systems
have been fully rigorously proved for quite general quantum spin systems in Ref.~\cite{BHM:stab,BH:stab}.

A more ambitious goal for robust quantum information processing is to genuinely store and manipulate quantum information
at nonzero temperature $T > 0$. Whether this is physically possible is the topic of the discussion on `self-correcting'
quantum memories~\cite{Dennis:2001,bacon:mem,BT:mem}, which
do not need to be continuously error-corrected as the standard theory
of quantum fault-tolerance prescribes \cite{AGP:ft}. The question of self-correction goes under alternative guises such as the question of thermal stability \cite{CLBT:mem,AHHH:4d} or thermal fragility \cite{Nussinov2008} of quantum memories or the
persistence of topological order at finite-temperature \cite{CC:toric,iblisdir+:mem,AFH:toric}.


In understanding aspects of topological order or its possible extension to finite temperature, it is important to study physical `toy' models such as the well-known surface code family or general quantum error-correcting codes with geometrically-local generators. These toy models both teach us what may be possible at the conceptual level as well as pose an interesting challenge to engineer these Hamiltonians at the physical level.

In this paper we introduce a class of toy models that can be viewed as extensions of Kitaev's 1D model of unpaired Majorana fermions in quantum wires~\cite{kitaev:unpair} to higher spatial dimensions and to interacting fermions.
These toy models which we call {\em Majorana fermion codes} can be described by fermionic term-wise commuting Hamiltonians composed of geometrically-local interactions on a $D$-dimensional lattice. The purpose of Majorana fermion codes
is to protect quantum information against low-weight fermionic errors, i.e., operators acting only on sufficiently small
subsets of fermionic modes. One distinction between fermionic systems and the systems composed of qubits or spins
is the presence of superselection rules. In particular, if a fermionic system interacts with a bosonic environment,
conservation of the parity of the total number of fermions restricts the set of physically realizable errors to the so-called even  fermionic operators.


One question addressed in the present paper is whether the superselection rules permit more robust storage of quantum
information based on Majorana fermion codes as compared with qubit stabilizer codes under the same geometric locality constraints.
We partially answer this question in the negative by generalizing the no-go theorem for quantum self-correction based on 2D stabilizer
codes~\cite{BT:mem} to Majorana fermion codes.

On the positive side, we construct interesting 2D generalizations of Kitaev's model of unpaired Majorana fermions in quantum wires~\cite{kitaev:unpair}. Specifically, we construct a family of Majorana fermion codes encoding one qubit into a 2D lattice of fermions
folded into a cylinder. The corresponding Hamiltonian has exactly two zero-energy Majorana modes,
i.e., odd fermionic  operators $\bar{C}_0,\bar{C}_1$ supported on the opposite edges
of the lattice, commuting with the Hamiltonian, and anti-commuting with each other.
In coding theory language, $\bar{C}_0$ and $\bar{C}_1$ are the logical operators of the code.
The main advantage of  the 2D model is that
the logical operators $\bar{C}_0,\bar{C}_1$ have a macroscopic weight proportional to the radius of the cylinder.
It endows the encoded qubit with an extra degree of  protection
related to the macroscopic distance of the code and which does not rely on superselection rules.
By varying the radius and the length of the cylinder one can combine the two types of protection in  a
controllable way.

The additional protection by the code distance is completely analogous to the protection of quantum information
encoded into the ground state of topologically ordered systems as discussed in~\cite{kitaev:anyons,Wen90,BHM:stab,BH:stab}.
Such protection might be necessary for example if one tries to build a multi-qubit register composed of 1D quantum wires
with unpaired Majorana modes. In this case a perturbation can couple the unpaired modes that belong to adjacent wires
without violating the superselection rules and hence an additional protection using a Majorana fermion code
with a large distance could be helpful.  In addition, the superselection rule prohibiting odd error operators is not likely to be
completely rigorous, for instance, if the environment supports gapless fermionic modes that can couple to the system, or,
when a single unpaired fermion could jump from the superconductor onto the topological insulator
(although such processes are energetically suppressed by a gap at low-enough temperature).


Finally, we argue that Majorana fermion codes can be used as a tool to prove new facts about qubit stabilizer codes.
In particular, we prove that any stabilizer code can be locally mapped onto a weakly self-dual Calderbank-Shor-Steane (CSS) code. The mapping preserves all parameters of the code such as the number of physical and logical qubits,
the distance, and locality of the generators up to a constant factor.


An open problem which is not addressed in our paper is by what physical means and mechanisms (interacting) Majorana fermion codes can be realized. One expects that, similar as for spin-systems, such models could emerge as effective many-body Hamiltonians for interacting fermion systems which are treated with a perturbative or renormalization-group flow analysis. There is ample physical evidence that 2D strongly-correlated electron systems support topological order; the question is whether the Majorana fermion code framework can help in understanding how such topological order emerges from the basic physical interactions.

The paper is organized as follows. Section~\ref{sec:why} reviews Kitaev's 1D model of unpaired Majorana fermions and provides some motivation behind the present work. Section~\ref{sec:defnot}
introduces notations and necessary facts from the theory of stabilizer codes.
A formal definition of Majorana fermion codes is given in Section~\ref{sec:mcdef}.
The mappings between stabilizer codes and Majorana fermion codes are described in Section~\ref{sec:cm}.
This section also proves the equivalence between general stabilizer codes and weakly self-dual CSS codes.
In Section~\ref{sec:majcode} we discuss quantum error correction in the presence of superselection rules. The generalization of Kitaev's 1D model with unpaired Majorana
fermions is described in Section~\ref{sec:mcc}.
We give more examples of Majorana fermion codes that possess odd logical operators in Section~\ref{sec:odd}.

\section{Why Majorana fermion codes?}
\label{sec:why}

In Ref. \cite{kitaev:unpair} Kitaev considered the toy Hamiltonian of a
1D chain of spinless fermions interacting with a superconductor.
The interaction with the superconductor allows for the creation and
annihilation of pairs of fermions so that the total Hamiltonian of
the system preserves only the parity of the number of fermions. Any Hamiltonian involving spin or spinless fermions can be written in terms of Majorana fermion operators by taking
$a_k=\frac{1}{2}(c_{2k-1}+i c_{2k})$ and
$a_k^{\dagger}=\frac{1}{2}(c_{2k-1}-i c_{2k})$. The Hermitian
Majorana operators obey the relations
\[
 c_i c_j+c_j c_i=2\delta_{ij}\,I.
\]
For a particular choice of couplings, Kitaev's Hamiltonian on $L$ spinless fermions, hence $2L$ Majorana fermions, reads
\be
\label{qwire}
H=i\sum_{j=1}^{L-1} c_{2j} c_{2j+1},
\ee
i.e. all Majorana modes are `paired', except the first mode $c_1$ and the last $c_{2L}$.
Thus this quadratic fermion Hamiltonian commutes with the operators $c_1$ and $c_{2L}$ which we identify as the logical operators of a protected qubit. Alternatively, it can be said that the symmetry gives rise to the presence of a pair of zero-energy Majorana boundary modes which lead to degeneracy at the Fermi-level.
Recently, there have been various proposals to realize such 1D model with zero-energy Majorana fermions, for example at the boundary between a topological insulator and a superconductor \cite{FK:topins} or in a semiconducting heterostructure \cite{LSDS:majferm}.

Ground states of $H$  are $-1$ eigenvectors for every
term $ic_{2j} c_{2j+1}$ in Eq.~(\ref{qwire}). The two-fold degenerate
ground subspace   has a basis
$\ket{\overline{0}}$  and $\ket{\overline{1}}$  which satisfies
\[
 \quad ic_1 c_{2L}  \ket{\overline{0}} = \ket{\overline{0}} \quad \mbox{and} \quad
\quad ic_1 c_{2L}  \ket{\overline{1}} = -\ket{\overline{1}}.
\]
The logical Pauli operators for a qubit encoded into $\ket{\overline{0}}$  and $\ket{\overline{1}}$  can be chosen
as $\bar{X}=c_1$, $\bar{Y}=-c_{2L}$, and $\bar{Z}=ic_1 c_{2L}$.
Note that two of the logical operators $c_1$ and $c_{2L}$ are of {\em odd} weight and hence would require the coherent creation/annihilation of a single fermion which is prohibited by superselection.
An essential part of the model is that the only even-weight  logical operator  $c_1 c_{2L}$ is very non-local. It is natural to assume that elementary perturbations to the Hamiltonian and errors can be represented by local even weight Majorana fermion operators. Hence in a perturbative analysis such as the Schrieffer-Wolf perturbation theory, the first contributions that split the energy degeneracy between
$\ket{\overline{0}}$ and $\ket{\overline{1}}$ are expected to occur in $O(L)$th order implying that the splitting in degeneracy between $\ket{\overline{0}}$ between $\ket{\overline{1}}$ is
exponentially small in $L$.

The spectrum and properties of Kitaev's model --as any other quadratic fermion Hamiltonian in the theory
of topological insulators-- are efficiently computable and quantum circuits which employ only non-interacting
fermion Hamiltonians and simple fermionic measurements are efficiently simulatable classically, see \cite{TD:fermions}. Hence, if we are serious about using fermionic systems to robustly store and manipulate quantum information (see e.g. \cite{blueprint:topo}), we will
need some source of interaction to obtain quantum universality (see e.g. \cite{BK:fermcode}).
Let us mention that generalizations of the Hamiltonian Eq.~(\ref{qwire}) to interacting fermions have been recently considered by Fidkowski and Kitaev~\cite{FK09}  to study the effect of interactions on the classification of 1D topological insulators.

The toy model, Eq.~(\ref{qwire}), demonstrates that fermionic parity conservation provides an alternative protection mechanism for the encoded qubit unrelated to topological quantum order.
It is therefore natural to ask whether topological protection can be combined with protection based on superselection rules in an advantageous way. The Majorana fermion codes introduced in the present paper provide a natural framework in which such a hybrid protection can be studied.



\section{Definitions and notations}
\label{sec:defnot}

We review a few standard definitions and notations which are used in this paper.
Let $X_i$,$Y_i$,$Z_i$ represent the three Pauli matrices on qubit $i$. Let $\calP_n=\la iI,X_1,Z_1,\ldots,X_n,Z_n\ra$ be the Pauli group on $n$ qubits generated by single-qubit Pauli operators and the phase factors $\pm1 $, $\pm i$. The support of a Pauli operator $P$, $\supp(P)$ is the set of qubits on which it acts non-trivially. The size of the support, $|\supp(P)|$, is also sometimes called the weight of $P$, denoted as $|P|$.

A pair of logical operators for an encoded qubit is denoted as $(\bar{X},\bar{Z})$ and $\bar{Y}=i\bar{X} \bar{Z}$.
A stabilizer code is determined by its {\em stabilizer group} $\calS\subseteq \calP_n$
which is an Abelian subgroup of $\calP_n$.
To preclude $\calS$ from containing non-trivial phase factors
one usually adds a requirement $-I\notin \calS$.
The set of Pauli operators $P\in \calP_n$ that commute with all elements
of $\calS$ is called the {\em centralizer} of $\calS$ and is denoted
as $\calC(\calS)$.
If the stabilizer group is generated by $n-k$ independent generators, then the centralizer is generated by $n+k$ independent generators.
Logical operators of a stabilizer code $\calS$ are elements of $\calC(\calS)$ which
are not in $\calS$. One can always choose a set of $2k$ logical Pauli operators
$\bar{X}_1,\bar{Z}_1, \ldots,
\bar{X}_k,\bar{Z}_k\in \calC(\calS)\backslash \calS$
obeying the usual Pauli commutation relations.
Note that $\calC(\calS)= \langle i, \calS,\bar{X}_1,\bar{Z}_1, \ldots, \bar{X}_k,\bar{Z}_k\rangle$.
Given a stabilizer code $\calS$, its codespace is spanned by all $n$-qubit states
invariant under the action of $\calS$. In this case the codespace is isomorphic to a space of $k$ qubits, the so called {\em logical qubits}.

The distance $d$ of a stabilizer code is defined as the minimum weight of a logical operator, i.e.
\be
d=\min_{P \in {\cal C}({\cal S})\backslash {\cal S}} |P|.
\ee
A stabilizer code which encodes $k$ logical qubits into $n$ physical qubits and has distance $d$ is denoted as a $[[n,k,d]]$ code.
We shall be interested in geometrically-local stabilizer codes and similarly local Majorana fermion codes.
For such codes, qubits occupy sites of a $D$-dimensional lattice (or, more generally, some
graph equipped with a metric) and
the stabilizer group has a set of geometrically-local generators,
$\calS=\la S_1,\ldots,S_m\ra$, that is, the support of any generator $S_i$
has diameter at most $r=O(1)$ with respect to the lattice geometry~\footnote{Here and below
the notation $f=O(g)$ refers to the limit $n\to \infty$. In particular,
$r=O(1)$ means that $r$ is upper bounded by a constant independent of $n$.}.

With a geometrically-local stabilizer code ${\cal S}$ we can associate a Hamiltonian, e.g. $H_{\cal S}=-\sum_i S_i$ where $S_i$ is an (over)complete set of stabilizer generators for ${\cal S}$. Then the ground-space of the Hamiltonian corresponds to the codespace and one may consider properties of such physical system at zero or non-zero temperature $T$.

Calderbank-Shor-Steane (CSS) codes are a particular subclass of stabilizer codes
 for which the stabilizer group $\calS$ can be represented as a product of two subgroups,
$\calS=\calS(X)\cdot \calS(Z)$, that
contain only $X$-type and $Z$-type Pauli operators respectively.
Any CSS code can be specified by a pair  of classical linear codes $C_X,C_Z\subseteq \{0,1\}^n$
such that
\[
\calS(X)=\{ P=\prod_{i=1}^n X_i^{x_i}\, : \,   (x_1,\ldots,x_n)\in C_X\},
\]
and
\[
\calS(Z)=\{ P=\prod_{i=1}^n Z_i^{z_i}\, : \,  (z_1,\ldots,z_n)\in C_Z\}.
\]
Commutativity between $\calS(X)$ and $\calS(Z)$ is equivalent to the mutual orthogonality of the
codes $C_X,C_Z$, that is, one must have $\sum_{i=1}^n x_i z_i =0 \pmod{2}$ for all $x\in C_X$
and $z\in C_Z$.
The special subclass of CSS codes obeying $C_X=C_Z$ are called weakly self-dual CSS codes.
For such codes the stabilizer group is invariant under the exchange of $X$ and $Z$ operators on every qubit. In that case the code $C=C_X=C_Z$ must be a weakly self-dual classical code, that is,
$\sum_{i=1}^n x_i z_i=0\pmod{2}$ for all $x,z\in C$. For more background on CSS codes, see~\cite{book:nielsen&chuang}.

\section{Definition of Majorana fermion codes}
\label{sec:mcdef}


We define a Majorana fermion code as follows.
Let $c_1,c_2,\ldots,c_{2n}$ be the $2n$ Majorana modes, i.e., operators
obeying commutation rules
\[
c_u c_v+ c_v c_u = 2\delta_{u,v}\, I, \quad c_u^\dag=c_u.
\]
The total number of Majorana modes ($2n$) is always even,
because these modes are obtained from $n$ original fermionic modes.
The single-mode operators $c_1,\ldots,c_{2n}$, together with the phase factor $i$, generate a group
of Majorana operators $\mjr{2n}$.
Any element of the group  $\mjr{2n}$ can be represented as
$\eta\,  c_A$ where $A\subseteq \{1,\ldots,2n\}$ is some subset of modes,
\be
\label{c_A}
c_A=\prod_{u\in A} c_u,
\ee
and $\eta\in \{\pm 1, \pm i\}$ is a phase factor.
The subset $A$ is called the support of $c_A$.
Here and below we
use a standard ordering of the product of single-mode operators
$c_u$, meaning that the indices $u$ increase from the left to the right.
We define the weight of a Majorana operator as the number of modes
in its support, that is, $|c_A|=|A|$. A Majorana operator
is called even (odd) iff its support has even (odd) size.

For two arbitrary supports $A$ and $B$ we have
\bea
c_A \, c_B =(-1)^{|A| \cdot |B|+|A\cap B|}\, c_B \, c_A.
\label{comm:maj}
\eea
When either $c_A$ or $c_B$ is even, the commutation relation only depends
on the parity of overlap $|A \cap B|$. In particular, when regions
$A$ and $B$ do not overlap, i.e. $|A \cap B|=0$, then  Majorana operators commute,
as is trivially the case for Pauli operators on non-overlapping supports.
However, when $c_A$ and $c_B$ are both odd and their supports are non-overlapping, then
$c_A$ and $c_B$ anti-commute.

One can always map the Majorana modes onto Pauli operators on
$n$ qubits using  the Jordan-Wigner transformation $\Upsilon\, : \, \mjr{2n}\to \calP_n$ defined as
$\Upsilon(c_{2i-1})=Z_1 Z_2 \ldots Z_{i-1} X_i$ and
$\Upsilon(c_{2i})=Z_1 \ldots Z_{i-1} Y_{i}$, $i=1,\ldots,n$.
Accordingly, the Hilbert space $\calF_n$ describing  $2n$ Majorana modes
can be identified with the Hilbert space of $n$ qubits.
 For $D> 1$ systems, the
Jordan-Wigner transformation generically maps local Majorana operators onto
non-local Pauli operators.
This is one of the reasons that local Majorana fermion codes defined below
may exhibit different properties than local stabilizer codes (see e.g. \cite{BK:fermcode,VC:fermionspin} for means to make this mapping local by introducing additional modes).

A Majorana fermion code is determined by its stabilizer group $\smaj \subseteq \mjr{2n}$
which must obey two conditions:
\begin{itemize}
\item $\smaj$ is an Abelian group not containing $-I$.
\item All elements of $\smaj$ have even weight.
\end{itemize}
The second condition guarantees that stabilizer operators preserve the parity of the number of fermions
in the system, and thus any element of  $\smaj$ is a physically realizable operation.
Given any pair of stabilizer operators proportional to $c_A$  and $c_B$, the commutativity $c_Ac_B=c_B c_A$
and the even-weight condition imply that the overlap $|A\cap B|$ must be even, see Eq.~(\ref{comm:maj}).

The set of Majorana operators $P\in \mjr{2n}$ that commute with all elements
of  $\smaj$ is called the {\em centralizer} of $\smaj$ and is denoted
as $\calC(\smaj)$. Note that the centralizer may contain both even and odd Majorana
operators. Logical operators of a Majorana fermion code  $\smaj$ are elements of $\calC(\smaj)$ which
are not in $\smaj$. If $\smaj$ is generated by $n-k$ independent generators, then $\calC(\smaj)$ is generated by $n+k$ independent generators. One can always choose a set of $2k$ logical Pauli operators
$\bar{X}_1,\bar{Z}_1, \ldots,
\bar{X}_k,\bar{Z}_k\in \calC(\smaj)\backslash \smaj$
obeying the usual Pauli commutation relations. Note that $\calC(\smaj)= \langle i, \smaj,\bar{X}_1,\bar{Z}_1, \ldots, \bar{X}_k,\bar{Z}_k\rangle$.
In general the set of logical operators may contain both even and odd Majorana operators.
We give necessary and sufficient conditions for a code to contain odd logical operators in Section~\ref{sec:cm},
see Proposition~\ref{prop:odd}.
The codespace of a Majorana fermion code $\smaj$ is the linear subspace of $\calF_n$ spanned by all states
invariant under the action of $\smaj$.

From now on, we shall be interested in geometrically-local Majorana fermion codes.
For such codes Majorana modes $c_u$ occupy sites of some $D$-dimensional lattice
$\Lambda$ and the stabilizer group $\smaj$ has a set of geometrically-local generators,
$\smaj =\la S_1,\ldots,S_m\ra$, that is, the support of any generator $S_i$
has diameter at most $r=O(1)$.
The codespace of $\smaj$ coincides with the ground subspace of a Hamiltonian
$H=-\sum_i S_i$ that involves only geometrically-local interactions among the Majorana modes.

The simplest example of a local  Majorana fermion code is the one describing Kitaev's
1D model of Eq.~(\ref{qwire}).
The corresponding stabilizer group is
\be
\label{Kitaev_code}
\smaj =\langle i c_{2}c_{3},\ldots, i c_{2n-2}c_{2n-1} \rangle,
\ee
while the logical operators of the code can be chosen as $\bar{X}=c_1$ and $\bar{Z}=c_{2n}$.
We describe a 2D generalization of this code in Section~\ref{sec:mcc} and give some other
examples of local Majorana fermion codes in Section~\ref{sec:odd}.

We can define the distance of a Majorana fermion code similar to the distance of  stabilizer codes, i.e.
as the minimum weight of logical operators,
\be
\label{dmaj}
d=\min_{C \in {\cal C}(\smaj)\backslash \smaj} |C|.
\ee
According to this definition, a code with distance $d$ is able to detect any error affecting less than $d$
Majorana modes, i.e., any operator $c_A$ with $|A|<d$ is a detectable error.

The notion of a code's distance does not completely
capture all aspects of stability that Majorana fermion codes can offer since it treats  even and odd logical operators
on the same footing. However, if the system is closed or interacts with a bosonic environment,
all physically realizable perturbations and error operators
must preserve fermionic parity and thus must be even.
In order to measure the degree of protection based on the superselection rules, let us
introduce an additional parameter $l_{even}$ defined as
the minimum diameter of a region that can support an even  logical operator,
\be
\label{lmaj}
l_{\rm even}=\min_{\ba{c} \scriptstyle C \in {\cal C}(\smaj)\backslash \smaj \\ \scriptstyle |C| =0 \pmod{2}\\ \ea } \mathrm{diam}(\supp{(C)}).
\ee
As far as zero-temperature stability  is concerned, the parameter $l_{\rm even}$ determines
the smallest order of perturbation theory at which the ground state
degeneracy of the code Hamiltonian $H=-\sum_i S_i$ can be lifted by a  perturbation that involves
only even geometrically local operators. Hence the parameters $d$ and $l_{even}$ capture two independent
mechanisms of protection: topological protection by the code distance and protection based on the superselection rules
respectively.

Consider as an example the code defined in Eq.~(\ref{Kitaev_code}).
Obviously, it has distance $d=1$.
Meanwhile, any even logical operator must include both $c_1$, $c_{2n}$, and therefore
$l_{\rm even}=2n$ (the lattice size).
More generally, using Lemma~\ref{fermdoub} from Section~\ref{sec:cm} and the fact (see~\cite{BT:mem}) that any 1D stabilizer code has distance $O(1)$, one can easily show that the  distance of any 1D Majorana fermion code is $O(1)$.

It is important to note that the minimum weight logical operator $C$ in Eq.~(\ref{dmaj})
always has a connected support,
that is, one cannot decompose $C$ as $C=c_Ac_B$ where
the separation between $A$ and $B$ is larger than $r$ (the largest diameter of
the generators of $\smaj$). Indeed, in this case both $c_A$ and $c_B$ would
individually commute with $\smaj$. Hence $c_A$ or $c_B$
would be a logical operator which contradicts the minimality of $C$.
On the other hand, the minimum weight even logical operator may have highly disconnected support
as the code in Eq.~(\ref{Kitaev_code}) demonstrates.

A simple argument shows that any Majorana fermion code has even logical operators. Indeed, if $\bar{X},\bar{Y},\bar{Z}$ are
logical Pauli operators for some encoded qubit then the identity $\bar{X} \bar{Y} \bar{Z} \propto I$
implies that either all $\bar{X},\bar{Y},\bar{Z}$ are even, or two of then
are odd and the third one is even.
We describe properties of
even logical operators for general $D$-dimensional Majorana fermion codes in
 Section~\ref{sec:majcode}.

As mentioned earlier, the superselection rule prohibiting odd error operators is not likely to be completely rigorous. For instance, one might be interested in constructions of Majorana fermion codes in which both $d$ and $l_{\rm even}$ can be made arbitrarily large by increasing lattice dimensions, see Section~\ref{sec:mcc},\ref{sec:odd}.

Before we continue, let us make a few remarks about general Majorana fermion codes based on {\em non-interacting} fermions. We would like to make the point that going to higher spatial dimensions $D>1$ does, in one aspect, not lead to fundamentally different behavior as compared to the 1D Kitaev's model of unpaired Majorana fermions~\cite{kitaev:unpair}. More specifically, the Bogoliubov transformation allows one to transform any non-interacting Majorana fermion Hamiltonian $H=i\sum_{k \neq l} \alpha_{kl} c_k c_l$ into a canonical form in which some subset of Majorana modes is unpaired (i.e. these modes do not enter into the Hamiltonian). Since each unpaired mode is a linear combination of the original Majorana operators $c_k$, the ground subspace of $H$ can be regarded as a quantum code with distance $d=1$. Hence non-interacting models can only offer protection based on superselection rules similar to what the 1D model of Ref.~\cite{kitaev:unpair} achieves. On the other hand, unlike in 1D, in 2D non-interacting fermion systems with unpaired Majorana modes, one can imagine adiabatically changing the Hamiltonian (or 'deforming' the quantum code \cite{BMD:deform}) to move localized unpaired Majorana modes around and enact some (but not all) logical gates by braiding.

\section{Code mappings}
\label{sec:cm}
In this Section we describe inter-conversions between three classes of codes:
(i) qubit stabilizer codes, (ii) Majorana fermion codes and (iii)  weakly self-dual CSS codes.
\begin{lemma}[Kitaev \cite{kitaev:anyon_pert}]
Every $[[n,k,d]]$ qubit stabilizer code $\calS$ can be mapped onto a Majorana fermion code $\smaj$ on $4n$ modes encoding $k$ logical qubits with distance $2d$.
\label{stabmaj}
\end{lemma}
For completeness, we give the mapping:

{\bf Proof}:
With every qubit $j$, we associate four Majorana fermion modes, $b_j^{x,y,z}$ and $c_j$ and hence we have a total of $4n$ Majorana fermions. In the $n-k$ independent stabilizer generators of ${\calS}$, we replace the local Pauli operators by
$X_j=i b_j^x c_j$, $Y_j=ib_j^y c_j$, $Z_j=i b_j^z c_j$. In addition, for each qubit $j$ we add a stabilizer $D_j=b_j^x b_j^y b_j^z c_j$ to $\smaj$ (on the subspace for which $D_j=+1$ we have
$X_j Y_j Z_j=iI$). Thus the Majorana fermion code $\smaj$ is generated by $2n-k$ independent generators, and therefore it encodes $k$ logical qubits.
Each logical operator of the stabilizer code corresponds to a logical operator of the Majorana code. Also, since a logical operator of the Majorana fermion code has to commute with each $D_j$, it must contain an even number of the set $\{ b_j^x, b_j^y, b_j^z, c_j\}$, and therefore it corresponds to a logical operator of $\calS$. Since every Pauli operator corresponds to a weight-2 Majorana operator, the distance of the Majorana fermion code is twice the distance of the stabilizer code. $\Box$.

Note that by this mapping every operator in the stabilizer $\smaj$ and logical operator in ${\calC}(\smaj)$ will have {\em even} weight. In addition, qubit errors get mapped onto even Majorana operators.

\begin{lemma}[Doubling]
With every Majorana fermion code $\smaj$ on $2n$ Majorana modes which encodes $k$ logical qubits and has distance $d$, we can associate a $[[2n,2k,d]]$ weakly self-dual CSS code.
\label{fermdoub}
\end{lemma}
Let us first illustrate the idea of the doubling map using the simplest Majorana code
with $4$ Majorana modes and a single generator $\smaj=\la c_1 c_2 c_3 c_4\ra$.
Clearly this code has $k=1$ logical qubit with logical operators
$\bar{X}_1=ic_1 c_2$ and $\bar{Z}_1=ic_1 c_3$. One can easily check that
$\smaj$ has distance $d=2$.
The doubled version of $\smaj$ is a $[[4,2,2]]$ stabilizer code with a stabilizer group
$\calS=\la X_1 X_2 X_3 X_4, Z_1 Z_2 Z_3 Z_4\ra$ obtained by replacing
each operator $c_u$ either with $X_u$ or $Z_u$ (hence the number of generators
is doubled). The logical operators of $\calS$ can be chosen
as $\bar{X}_1=X_1 X_2$, $\bar{Z}_1=Z_1 Z_3$, $\bar{X}_2=X_1 X_3$, and $\bar{Z}_2=Z_1 Z_2$.

\noindent
{\bf Proof of Lemma~\ref{fermdoub}:}
Any operator  $P\in \mjr{2n}$ can be parameterized (up to a phase factor)
by a binary string $x\in \{0,1\}^{2n}$ such that
multiplication in $\mjr{2n}$ corresponds to addition of binary strings modulo two.
Specifically, one sets $x_u=1$ if $u$ belongs to the support of $P$ and $x_u=0$
otherwise. Let $\phi\, : \, \mjr{2n} \to \{0,1\}^{2n}$ be the corresponding mapping.
Consider a classical code
\[
C=\phi(\smaj)\subset \{0,1\}^{2n}.
\]
Note that $\dim{(C)}=n-k$,
since $\smaj$ has $n-k$ independent generators.
Furthermore,  since $\smaj$ is an Abelian group
containing only even operators, the  supports of any elements $P,Q\in \smaj$
must have even overlap. Hence $C$ is a weakly self-dual classical code, that is,
$\sum_{i=1}^{2n} x_i y_i=0\pmod{2}$ for any $x,y\in C$, or, equivalently,
\[
C\subseteq C^\perp.
\]
Let $\calS=\calS(X)\cdot \calS(Z)$ be the weakly self-dual CSS code constructed from  $C$ as explained in
Section~\ref{sec:defnot}. By construction, the  code $\calS$ has $2n$ qubits,
$n-k$ independent generators of $X$-type, and $n-k$ independent generators of $Z$-type.
Hence $\calS$ encodes $2k$ qubits.
Note that each generator of $\smaj$ gives rise to a pair of generators in $\calS$:
the one obtained by replacing each single-mode operator $c_j$ with $X_j$, and
the one obtained by replacing each single-mode operator $c_j$ with $Z_j$.

Consider a minimum-weight logical operator in the code ${\cal S}$; w.l.o.g. it
is either a product of $X$ or a product of $Z$s (but not of both). When we replace each Pauli
$X_i$ (or $Z_i$) by a Majorana operator $c_i$, we obtain a logical operator for the Majorana code.
Vice versa, every logical Majorana operator gives rise to a pair of logical operators for
the stabilizer code ${\cal S}$. Hence the distances of these codes are identical. $\Box$.

Combining the two lemmas, we get the following useful fact.
\begin{corol}
Any $[[n,k,d]]$ stabilizer code can be mapped onto a $[[4n,2k,2d]]$ weakly self-dual CSS code.
This mapping preserves geometric locality of a code up to a constant factor.
\end{corol}
 This result thus shows that in order to derive distance bounds for, say, geometrically-local codes, one only needs to prove such bounds for weakly self-dual CSS codes and the scaling of rates and relative overhead can be determined by considering only weakly self-dual CSS codes. In addition, the code mappings allow one to show that the partition function of a Hamiltonian associated with a stabilizer code can be expressed as the partition function of a classical Ising (${\bf Z}_2$) gauge theory. Since the mapping preserves the locality of errors, it is also the physics which is preserved. Hence the properties of ${\bf Z}_2$-gauge models, the presence of a phase-transition or not, will be related to the question of thermal stability of any stabilizer code \cite{Kogut:RMP}. The weakly self-dual character of the CSS code, or the fact that the Hamiltonian of the ${\bf Z}_2$-gauge model has only terms with even overlap, is crucial. For example, it is well-known that there exists a 3D Ising gauge model \cite{Wegner71} (in fact, this model is the $Z$-part (i.e. the subgroup ${\cal S}(Z)$ of the stabilizer ${\cal S}$) of the 3D surface code which was shown to be thermally stable against $X$-errors \cite{CC:3Dtoric}) which has a phase-transition at a non-zero temperature $T_c$. However, this Ising gauge model does not have the property that all terms have even overlap. Hence this model is not directly pertinent to the thermal stability of 3D stabilizer codes for which one needs macroscopic energy barriers against both against $X$- {\em and} $Z$-error excitations.


\section{Properties of Majorana fermion codes}
\label{sec:majcode}


In the previous section we have seen that any local Majorana fermion code can be mapped to a local weakly self-dual CSS code
without changing parameters of the code in any significant way, see Lemma~\ref{fermdoub}.
Therefore one directly apply any distance upper bounds obtained for local stabilizer codes in~\cite{BT:mem,BPT:tradeoff} to obtain analogous upper bounds on local Majorana fermion codes. However, one might expect that a Majorana fermion code may offer an additional degree of protection resulting from conservation of the fermionic parity. Such additional protection may only manifest itself for Majorana fermion codes that possess odd logical operators,
since for such codes at least some subset of the logical operators is protected by the superselection rules, see Section~\ref{sec:mcdef}.
Hence, the first question  we address in this section is under what conditions
a Majorana fermion code has at least one odd logical operator.
Let us define an operator $\Call$ measuring the parity of the total number of fermions,
\be
\label{Call}
\Call = i^n c_1 c_2 \cdots c_{2n-1} c_{2n}.
\ee
\begin{prop}
\label{prop:odd}
A Majorana fermion code $\smaj$
has at least one odd logical operator iff $\Call \notin \pm \smaj$.
\end{prop}

{\bf Proof:}
Indeed, suppose $\Call \in \pm \smaj$.
Then any logical operator $P \in \calC(\smaj) \backslash \smaj$ must have even weight, since it has to commute with
$\Call$. Suppose now that $\Call \notin \pm \smaj$.
Since the support of $\Call$ has even overlap with the support of any element of $\smaj$, we conclude that
$\Call$ is a logical operator. Then there must exist another logical operator $P$ which anti-commutes with
$\Call$. But this is possible only if $P$ has odd weight. $\Box$



Let us point out that for any Majorana fermion code there is a choice of logical Pauli operators
such that at most one of them is odd. Indeed, if a code has at least one odd logical operator,
we can choose logical Pauli operators on the first encoded qubit as $\bar{Z}_1=\Call$
and $\bar{X}_1=P$, where $P$ is some odd logical operator, see Proposition~\ref{prop:odd}.
Since the logical Pauli operators $\bar{X}_i$, $\bar{Z}_i$ on the remaining qubit must commute with
$\bar{Z}_1$, they must be even.
It will be convenient to introduce a parameter $k_{\rm odd}\in \{0,1\}$
such that $k_{\rm odd}=1$ iff a code has at least one odd logical operator. (It is perhaps important to note that $k_{\rm odd}=1$ does not imply that at most one logical qubit has odd logical operators. In fact, one can show that there always exists a choice for the logical operators such that {\em all} logical qubits have logical $\bar{X}$ and $\bar{Z}$ operators which are odd weight.)

For 2D stabilizer codes, it has been proved in~\cite{BT:mem} that
one can always find logical operators which are supported on
 a strip of constant width, i.e., have string-like geometry.
Lemma~\ref{fermdoub} immediately shows that the same result holds for 2D Majorana fermion codes.
In particular, the distance of any 2D Majorana fermion code defined on a lattice of size $L\times L$
is at most $O(L)$.
However, as we mentioned earlier in the paper, for Majorana fermion codes with $k_{\rm odd}=1$
we have to focus only on the even logical operators while odd logical operators are prohibited by the
superselection rules.
For example, the results of~\cite{BT:mem} do not rule out the possibility that
a 2D Majorana fermion code may have logical operators $\bar{X},\bar{Y},\bar{Z}$ such that
$\bar{X}$ is even, $\bar{Y},\bar{Z}$ are odd, and
$\bar{X}$ has a plane-like geometry, that is, the minimum weight of  $\bar{X}$ is of order $n\sim L^2$. Such a code might behave similar to the classical 2D ferromagnetic Ising model
 in terms of its thermal stability.
Unfortunately, below we prove that 2D Majorana fermion codes do not behave like the 2D ferromagnetic Ising model;
more precisely we will show the following.

\begin{lemma}
Let $\smaj=\la S_1,\ldots,S_m\ra$ be a local Majorana fermion code defined on a  2D  lattice (with periodic or open boundary conditions) such that the support of any generator $S_a$ has diameter at most $r-1$ for some constant $r$.
Let $\Lambda=A_1 \cup A_2 \cup \ldots \cup A_t$ be a partition of the lattice into parallel disjoint strips of  width at least $r$.
Then one of the following (or both) is true:
\begin{enumerate}
\item There exists an even logical operator $\bar{C}$ supported on some strip $A_i$;
\item There exists a pair of odd logical operators $\bar{C}_i$, $\bar{C}_j$ supported on some
pair of strips $A_i, A_j$, $i\ne j$.
\end{enumerate}
\label{lem:upper}
\end{lemma}
Let us first comment on the implications of the lemma. Consider first the case (i).  In this case the
code has an even logical operator $\bar{C}$ whose support is confined to a rectangular region of size $r\times L$,
where $L$ is the lattice size. In other words, at least one even logical operator has string-like geometry.
Consider now the case~(ii). Since the logical operators $\bar{C}_i$, $\bar{C}_j$ are odd and have non-overlapping
supports, they must anti-commute. It implies that $\bar{C}_i$ and $\bar{C}_j$ are distinct logical operators,
that is, $\bar{C}_i \bar{C}_j \notin \smaj$. But then $\bar{C}_i \bar{C}_j$ is an even logical operator whose
support consists of two disjoint string-like regions.
This result suggests that 2D Majorana fermion codes cannot surpass 2D stabilizer codes in terms of their thermal stability by gaining additional protection based on the superselection rules. Indeed, as was pointed out by many authors~\cite{Dennis:2001,bacon:mem,kay:nonreliable,BT:mem,CLBT:mem}, the existence of string-like logical operators and the lack of string-tension rule out the possibility of quantum self-correction at a non-zero temperature. More in particular, a logical operator supported on two string-like regions, can be generated by a sequence of local even Majorana fermion operators such that for every state obtained in the sequence its energy is $O(1)$ above the ground-state energy. This argument shows that the energy barrier between logical states is $O(1)$, see~\cite{BT:mem}.


Note that Lemma~\ref{lem:upper} can be applied to 1D geometry as well by considering a 2D lattice
of size $L\times 1$. In this situation the strips $A_i$ become intervals of a 1D chain of length $r-1$.
Obviously, if a 1D code satisfies  case~(i) of Lemma~\ref{lem:upper}, it does not provide any protection
at all, since it has an even logical operator of constant weight and constant diameter.
On the other hand, a 1D code satisfying case~(ii) of   Lemma~\ref{lem:upper}
behaves similar to the Kitaev's 1D model, see Eq.~(\ref{qwire}). Indeed, such a code
has two constant-weight odd logical operators $\bar{C}_i$, $\bar{C}_j$
supported on some disjoint intervals $A_i,A_j$.
Clearly, the largest possible diameter of the corresponding even logical operator $\bar{C}_i \bar{C}_j$ is of order $L$.
Using the notation of Section~\ref{sec:mcdef},
any 1D Majorana code must obey
\be
\label{1Ddist}
d=O(1) \quad \mbox{and} \quad  l_{\rm even}=O(L).
\ee
In particular, it shows that Kitaev's 1D Majorana chain demonstrates the optimal behavior
even among the subclass of {\em interacting} fermionic models corresponding to Majorana fermion codes.

In Section~\ref{sec:mcc} we construct a 2D Majorana fermion code that
demonstrates the optimal behavior allowed by Lemma~\ref{lem:upper}.
This code has a single even logical operator and two odd logical operators
with weight of order $L$ located on the opposite boundaries of the lattice.

Lemma~\ref{lem:upper} can be straightforwardly generalized to any spatial dimension $D$
using the partition $\Lambda=A_1 \cup A_2 \cup \ldots \cup A_t$ into disjoint hyper-strips of width at least $r$, that is,
rectangles of size $s\times L\times \ldots \times L$, where $L$ is the lattice size and $s \geq r$.

In the rest of the section we prove Lemma~\ref{lem:upper}.

\noindent
{\bf Proof:}
Let us say that a subset of the lattice $M\subseteq \Lambda$ is {\em cleanable} iff
for any logical operator $\bar{D}\in \calC(\smaj)\backslash \smaj$ there exists a stabilizer $S\in \smaj$
such that $\bar{D}S$ acts trivially on $M$,
that is, $\supp{(\bar{D}S)}\cap M=\emptyset$.
Otherwise we shall say that $M$ is {\em uncleanable}.
We shall use the following simple fact.
\begin{prop}
\label{prop:cl}
 A subset $M\subseteq \Lambda$ is uncleanable iff there exists a logical operator $\bar{C}\in \calC(\smaj)\backslash \smaj$
supported on $M$.
\end{prop}
The proposition can be easily proved by applying the Cleaning Lemma from~\cite{BT:mem}
to the doubled stabilizer code constructed from $\smaj$ as described in Section~\ref{sec:cm}.
For the sake of completeness, we give a more direct proof of Proposition~\ref{prop:cl}
in Appendix~A.

Now consider two cases:
\begin{enumerate}
\item[\bf (1)] There are at least two uncleanable strips $A_i$, $A_j$, $i \not= j$;
\item[\bf (2)] There is at most one uncleanable strip $A_i$ .
\end{enumerate}
Consider first case (1). Let $\bar{C}_i$ and $\bar{C}_j$ be the logical operators of $\smaj$ supported on $A_i$ and $A_j$
which exist by Proposition~\ref{prop:cl}.
If at least one of $\bar{C}_i$, $\bar{C}_j$ is even, we arrive at case~(i) of the lemma.
If both $\bar{C}_i$, $\bar{C}_j$ are odd, we arrive at case~(ii).

Let us now consider case~(2).
We shall color the strips in black and white in the alternating order
such that the only uncleanable strip $A_i$ (if any) is black.
Then every white strip  is cleanable. Moreover, since the generators of $\smaj$
have diameter smaller than the width of a strip, the union of all white strips is also
cleanable. As was mentioned in Section~\ref{sec:mcdef}, we can always choose at least one
even logical operator $\bar{D}\in \calC(\smaj)\backslash \smaj$.
Let $S\in \smaj$ be the stabilizer that cleans $\bar{D}$ from the union of white strips.
Then  $\bar{C}:=S\bar{D}$ is an even logical operator of $\smaj$ that has support only on
black strips. Let $\bar{C}_i$ be the restriction of $\bar{C}$ onto a black strip $A_i$.
Note that for any $i$ the operator $\bar{C}_i$ is either a stabilizer or a logical operator of $\smaj$.
If there exists a black strip $A_i$ such that $\bar{C}_i$ is a logical operator with even weight, we arrive at case~(i)
of the lemma.
Otherwise, the number of black strips $A_i$ such that $\bar{C}_i$ is a logical operator with odd weight must be non-zero and even
(recall that the overall weight of $\bar{C}$ is even). Hence we can choose a pair of black strips $A_i$, $A_j$
such that $\bar{C}_i$ and $\bar{C}_j$ are logical operators with odd weight.
We arrive at case~(ii) of the lemma. $\Box$


\section{Majorana color code}
\label{sec:mcc}


In this Section we describe a fermionic version of the topological color codes introduced by Bombin and Martin-Delgado in~\cite{BMD:topo}. The color codes are weakly self-dual CSS codes with geometrically-local generators.
One can define a color code on any two-dimensional lattice or, more generally, on any surface graph which is $3$-valent and admits a $3$-coloring of its faces. For such graphs two faces share an \emph{even} number of vertices, as is easily checked. Given such a graph, the color code is defined by placing qubits at the vertices of the graph. The generators of the stabilizer group are associated with  faces of the lattice. Specifically, for every face $f$ one defines a pair of generators $S_f(X)$ and $S_f(Z)$ equal to the product of Pauli $X$'s and $Z$'s respectively over all qubits lying on the boundary of $f$.
The even overlap condition guarantees that all generators pairwise commute. One can show that the logical operators of the color code can be identified with homologically non-trivial loops on the lattice, see~\cite{BMD:topo} for details.

Lemma~\ref{fermdoub} allows one to identify any color code with a doubled Majorana fermion code. A simple example is the 2D color code on a hexagonal lattice with periodic boundary conditions (i.e. a torus). Such code encodes $4$ logical qubits, see~\cite{BMD:topo}.  The corresponding Majorana fermion code has a single Majorana mode $c_u$
at every site of the lattice and a single generator $C_f$ at every hexagon $f$.
The generator $C_f$ is proportional to the product of single-mode operators $c_u$ over all sites $u$ lying on the boundary of $f$. Note that the pair of generators $S_f(X),S_f(Z)$ can be obtained
by applying the doubling transformation of Lemma~\ref{fermdoub} to the generator $C_f$.
Hence the Majorana code with stabilizers $C_f$ encodes $2$ logical qubits.
 It is important to note that all 2D color codes discussed in
Ref.~\cite{BMD:topo} have only even-weight logical operators
(with the exception of the so-called triangular codes which we discuss in Section~\ref{sec:odd}).
 The absence of odd-weight logical operators in a color code implies the absence of odd logical operators in its fermionic version. Hence superselection rules do not play a role in their stability properties.

The formalism developed by Bombin and Martin-Delgado in~\cite{BMD:topo} employs $3$-coloring of the set of the faces of the lattice to classify the logical operators of the code. As we shall see, the global face $3$-coloring condition is too restrictive as it leaves many interesting color-type codes beyond the scope of the formalism. In particular, the
Majorana color codes that we describe below are defined on lattices that admit only a {\em local}  $3$-coloring, meaning that
any topologically trivial region of the lattice admits a face $3$-coloring but it cannot be extended to the entire lattice.
The peculiar feature of such codes is that they possess odd logical operators.

\begin{figure}
\centerline{
\includegraphics[height=4cm]{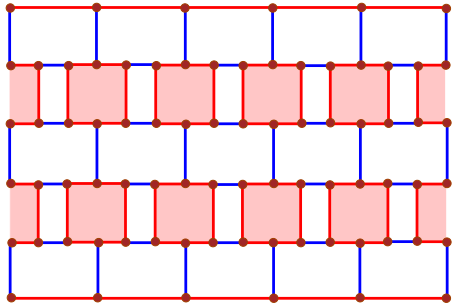}}
\caption{Example of a surface graph satisfying conditions (G1)-(G4).
The lattice has periodic boundary conditions along the horizontal axis
and open boundary conditions along the vertical axis.
The subset $F_0$ consists of $10$ faces (shaded hexagons).
The edges of $E_0$ and $E_1$ are shown using the blue and red color respectively.
The boundary components $\gamma_0,\gamma_1$ consist of $5$ vertices.}
\label{fig:example}
\end{figure}

Let $\Sigma=S^1\times [0,1]$ be a two-dimensional cylinder and $G\subseteq \Sigma$ be a graph embedded in $\Sigma$.
We shall assume that the graph $G$ induces a cellular decomposition of $\Sigma$, that is, the surface
$\Sigma$ can be decomposed into
a set of faces, edges, and vertices that we shall denote $F$, $E$, and $V$ respectively.
The boundary $\partial \Sigma$ consists of two cycles
$S^1\times \{0\}$ and $S^1\times \{1\}$.
In order to define a Majorana color code, we shall impose four conditions on the graph $G$:
\bea
\mbox{\bf (G1)} &\quad &  \mbox{The total number of vertices is even.} \nn \\
\mbox{\bf (G2)} &\quad & \mbox{Each vertex has degree $3$ (trivalent graph).} \nn \\
\mbox{\bf (G3)} &\quad &  \mbox{The boundary of any face has even length.} \nn \\
\mbox{\bf (G4)} &\quad & \mbox{The boundaries $S^1\times \{0\}$ and $S^1\times \{1\}$ have odd length.}  \nn
\eea
Given a face $f\in F$, let $V(f)\subseteq V$ be the set of all vertices that lie on the boundary of $f$.
We shall say that a face $f\in F$ is adjacent to a vertex $u\in V$ iff $u\in V(f)$, and we say that two faces are adjacent if they share a common edge.
The assumption that $G$ induces a cellular decomposition of $\Sigma$ together with condition (G2) imply
that any vertex $u\notin \partial \Sigma$ has exactly three adjacent faces and any vertex $u\in \partial \Sigma$
has exactly two adjacent faces.
This, together with (G4), implies via Proposition~\ref{prop:odd} that we must have odd logical operators.
We show a non-trivial example of a surface graph $G$ satisfying conditions (G1-G4) in Fig.~\ref{fig:example}.


Suppose each vertex $u\in V$ is occupied by a Majorana mode $c_u$.
  For any face $f\in F$ we define a {\em face operator}
\be
C_f= \prod_{u\in V(f)}\, c_u.
\ee
Conditions (G2) and (G3) imply that all face operators have even weight
and any pair of face operators commute with each other. Using the standard stabilizer formalism one can show that there exists a choice of phase factors $\eta_f\in \{\pm 1, \pm i\}$, $f\in F$, such that operators  $\eta_f  \,  C_f$
generate an Abelian subgroup $\smaj(G)\subseteq \mjr{2n}$ not containing minus identity, i.e.
\be
\smaj(G)=\la \eta_f \, C_f, \quad f\in F\ra.
\ee
is a Majorana fermion code. This code will be referred to as a {\em Majorana color code} associated with $G$.
One can also regard the codespace of $\smaj(G)$ as the ground subspace of
a fermionic local Hamiltonian
\be
\label{Hcolor}
H=-\sum_{f\in F} \eta_f\, C_f
\ee
(Although  the coefficients $\eta_f$ do not affect any parameters of
the code, one needs to have an explicit expression for $\eta_f$ to define the Hamiltonian model Eq.~(\ref{Hcolor}).
Let us mention that one can explicitly compute $\eta_f$
using the geometrical structures known as {\em Kasteleyn orientations} of a surface~\cite{Reshetikhin07,Reshetikhin08,
matchgates}).

We begin by describing the logical operators of the code.
Let $\gamma_0$ and $\gamma_1$ be the two boundary components of $\Sigma$, that is,
\be
\gamma_0=V\cap (S^1\times \{0\}) \quad \mbox{and} \quad \gamma_1=V\cap (S^1\times \{1\}).
\ee
Define operators
\be
\bar{C_\alpha}=\prod_{u\in \gamma_\alpha}\, c_u, \quad \alpha=0,1.
\ee
Note that for any face $f$, the set $V(f)$ has even overlap with $\gamma_\alpha$ since
one can regard $\gamma_\alpha$ as a boundary of an external face obtained by patching up the hole in $\Sigma$.
Therefore $\bar{C_\alpha}$ commutes with any face operator. On the other hand,
condition (G4) implies that $\bar{C_\alpha}$ has odd weight, and thus $\bar{C_\alpha} \notin \smaj(G)$.
We conclude that $\bar{C_\alpha}$ are logical operators of the code.
In addition, since $\bar{C_0}\bar{C_1}=-\bar{C_1}\bar{C_0}$, these are two independent logical operators. In other words, the Majorana color code encodes at least one qubit and the logical Pauli operators for this qubit can be chosen as
\be
\label{logP}
\bar{X}=\bar{C_0}, \quad \bar{Y}=\bar{C_1}, \quad \bar{Z}=-i\bar{C_0}\bar{C_1}.
\ee
The following lemma shows that this is the only logical qubit.
\begin{lemma}
\label{lemma:kmaj}
The Majorana color code has exactly one logical qubit, i.e., $k=1$.
\end{lemma}
{\bf Proof:}
Let us define a classical linear code $\calC\subseteq \{0,1\}^{|F|}$ whose codewords describe linear dependencies
among the face operators. Specifically, a binary string $x=\{x_f\}_{f\in F}$ is a codeword of $\calC$ iff
\be
\label{red1}
\prod_{f\in F} C_f^{x_f} \propto I.
\ee
Using the standard stabilizer formalism one can show that
\be
k=\frac{|V|}{2} - \dim{(\smaj(G))} = \frac{|V|}2 - |F| + \dim{(\calC)} = \dim{(\calC)},
\ee
where the last equality follows from the Euler formula
$|V|+|F|-|E|=0$ and the identity $3|V|=2|E|$.
Since we have already shown that $k\ge 1$, we know that Eq.~(\ref{red1}) has at least one
non-trivial solution, that is,
\be
\dim{(\calC)}\ge 1.
\ee

Let $u$ be a vertex in $\partial \Sigma$, and let $f$ and $g$ be the two adjacent faces. We claim that if we fix the values of $x_f$ and $x_g$ we uniquely determine a solution to Eq.~(\ref{red1}). Since we see from Eq.~(\ref{red1}) that $x_f \oplus x_g = 0$, we then proved that there are at most two solutions, and hence $\dim{\calC} \leq 1$.

To prove the claim, let $v \in V$ be another vertex of $G$, and let $\omega = (u_0 = u, u_1, \ldots, u_t = v)$ be any path connecting $u$ and $v$.  Suppose we have already set the value of $x$ on some pair of  faces $f_i$, $g_i$ adjacent to the vertex $u_i$ for some $i\ge 0$. Consider two cases.
{\em Case~1:} $u_{i}\notin \partial \Sigma$. Let $h_i$ be the third face adjacent to $u_i$. From  Eq.~(\ref{red1}) we infer that $x_{f_i}\oplus x_{g_i}\oplus x_{h_i}=0$ which uniquely sets $x_{h_i}$. Since two of the faces $f_i,g_i,h_i$ are adjacent to the edge $(u_i,u_{i+1})$, it sets the value of $x$ on some pair of faces adjacent to $u_{i+1}$.
{\em Case~2:} $u_i\in \partial \Sigma$. In this case $u_i$ has only two adjacent faces $f_i,g_i$. If $u_{i+1}\notin \partial \Sigma$, then both  faces $f_i,g_i$ are adjacent to $u_{i+1}$. If $u_{i+1}\in \partial \Sigma$ then only one of the faces  $f_i,g_i$ is adjacent to $u_{i+1}$, say the face $f_i$. Let $f_{i+1}=f_i$  and $g_{i+1}$ be the two faces adjacent to $u_{i+1}$. From Eq.~(\ref{red1}) we infer that $x_{f_{i+1}}\oplus x_{g_{i+1}}=0$ which sets the value of $x$ on the two faces adjacent to $u_{i+1}$. This actually even shows that $x$ has to have the same value for all faces along a common boundary of $\Sigma$.

Applying this argument inductively one sets the value of $x$ on some pair of faces adjacent to $v$ which sets the value of $x$ on all faces adjacent to $v$. Since any face is adjacent to some vertex, it shows that there is at most one way to extend $x_{f}$ and $x_{g}$   to a solution of Eq.~(\ref{red1}).
$\Box$

The unique non-trivial solution of Eq.~(\ref{red1}) constructed above
allows one to define subsets of faces $F_0=\{ f\in F\, : \, x_f=0\}$
and $F_1=\{f\in F\, : \, x_f=1\}$ such that any vertex has exactly two adjacent faces from $F_1$,
see Fig.~\ref{fig:example} in which the faces from $F_0$ are represented by shaded hexagons.
In other words, we have the following corollary.
\begin{corol}
\label{cor:F}
There exists a unique partition of the set of faces $F$ into disjoint subsets $F_0$ and $F_1$
such that each vertex has exactly two adjacent faces from $F_1$
and each vertex not lying on the boundary $\partial \Sigma$ has exactly one
adjacent face from $F_0$.
\end{corol}
Another interesting corollary of Lemma~\ref{lemma:kmaj} is that
the graph $G$ is not face $3$-colorable. Recall that a face $3$-coloring is
a mapping $c\, : \, F \to \{0,1,-1\}$ such that for any pair of adjacent
faces $f,g$ one has $c(f)\ne c(g)$.
\begin{corol}
\label{cor:3}
The graph $G$ does not permit face $3$-coloring.
\end{corol}
{\bf Proof:} Indeed, suppose such a $3$-coloring exists.
Then clearly all faces in $F_0$ must have the same color, say, $c(f)=0$ for all $f\in F_0$.
It implies that all faces adjacent to the boundary $\partial \Sigma$ must be colored by $\pm 1$.
However, since the boundary components have odd length, such a coloring does not exist.
$\Box$

One can use Corollary~\ref{cor:F} to define a family of even-weight logical operators
whose supports have geometry of a string connecting the two boundary components of $\Sigma$.
Indeed, define a subset of edges
\be
\label{E_0}
E_0=\{ e\in E\, : \quad \mbox{both faces adjacent to $e$ belong to $F_1$}\}.
\ee
Any edge $e\in E_0$ connects some pair of distinct faces in $F_0$,
or connects some face in $F_0$ with one of the two external faces $f_{{\rm ext},0}$, $f_{{\rm ext},1}$ obtained by patching
the holes in $\Sigma$ (see Fig.~\ref{fig:example} where
the edges from $E_0$ are shown in blue). Given any edge $(u,v)\in E_0$
connecting some pair of faces $f,g\in F_0$,
the operator $c_u c_v$ commutes with any face operator $C_f$, $f\in F_1$, and anticommutes with $C_f$ and $C_g$.
Hence we can construct logical operators associated with paths of edges in $E_0$
that connect the two external faces $f_{{\rm ext},0}$ and $f_{{\rm ext},1}$.
More specifically,  consider a graph $G^{(0)}$ with a set of vertices $F_0\cup f_{{\rm ext},0} \cup f_{\rm{ ext},1}$
and a set of edges $E_0$. Let $\gamma=(e_1,\ldots,e_m)$, $e_i\in E_0$,  be any path on $G^{(0)}$ connecting
$f_{{\rm ext},0}$ and $f_{{\rm ext},1}$. Then the operator
\be
\label{Cgamma}
\bar{C}_\gamma=\prod_{(u,v)\in \gamma} \, c_u c_v
\ee
commutes with all face operators $C_f$, $f\in F$.
On the other hand, $\bar{C}_\gamma$ anticommutes with $\bar{c}_0$ and $\bar{c}_1$
since it shares exactly one vertex with the boundaries $\gamma_0$ and $\gamma_1$.
We conclude that $\bar{C}_\gamma$ is the  logical operator $\bar{Z}\sim \bar{c}_0\bar{c}_1$, see Eq.~(\ref{logP}).

The graph $G^{(0)}$ defined  above allows one to construct a face $3$-coloring for any
topologically trivial region of the lattice. Indeed, as was pointed out above,
all faces $f\in F_0$ must have the same color, for instance, $c(f)=0$ for all $f\in F_0$. Then one needs
to color the faces  $f\in F_1$ using the colors $c(f)=\pm 1$ such that adjacent faces in $F_1$
have different colors. Recall that any pair of adjacent faces in $F_1$ can be identified with
some edge $e\in E_0$, see Eq.~(\ref{E_0}). Hence $G$ admits a face $3$-coloring iff the graph dual to $G^{(0)}$
admits a vertex $2$-coloring. Let us denote this dual graph $G^{(1)}$. It has the set of vertices $F_1$
and  the set of edges $E_0$. The set of faces of $G^{(1)}$ can be identified with $F_0$.
Since each face $f\in F_0$ has even-length boundary, any homologically trivial cycle in $G^{(1)}$
must have even length. Hence one can construct a vertex $2$-coloring of any subgraph $G^{(1)}$
that does not contain homologically non-trivial cycles.

Let us now bound the distance  of the Majorana color code focusing on the physically
relevant case when the generators of the code are {\em geometrically-local}.
We shall assume that $\Sigma$ is equipped with a metric such that the boundary components $S^1\times \{0\}$
and $S^1\times \{1\}$ have length $R$, while the
distance between them is $L$. We also assume that edges of $G$ have length at most $O(1)$
and any face consists of $O(1)$ edges.
Below we prove that the distance of the code grows linearly
with the smallest of the surface dimensions, namely,
\be
\label{mccd1}
d =\Omega(\min{(R,L)}).
\ee
As for the minimum diameter  of even logical operators, we prove the bound
\be
\label{mccd2}
l_{\rm even}=\Omega(L),
\ee
see Section~\ref{sec:mcdef} for notations.
The regime in which $R=O(1)$ while $L\gg 1$
can be regarded as protection by the superselection rules only
since in this regime the code behave similarly to the Kitaev's 1D model, see Eq.~(\ref{1Ddist}).
The regime in which both dimensions $R,L\gg 1$ are of the same order can be regarded
as protection by the code distance only, since in this regime both even and odd logical operators
are equally difficult to implement. In the intermediate regimes the code combines
both types of protection in a way that can be controlled by the choice of $R$ and $L$.

Let us now prove the bounds Eqs.~(\ref{mccd1},\ref{mccd2}).
We start from observing that any odd logical operator must have weight
 $\Omega(R)$.
Indeed,  any such logical operator $\bar{P}$ must anti-commute with even logical
operators $\bar{C}_\gamma$ constructed above,  see Eq.~(\ref{Cgamma}). Obviously, one can choose
$m$ pairwise disjoint paths $\gamma_1,\ldots,\gamma_m$ on the graph $G^{(0)}$ connecting the two external faces
where $m=\Omega(R)$. Then the support of $\bar{P}$ must have odd overlap
with each of the paths $\gamma_i$, $i=1,\ldots,m$, that is, the weight of $\bar{P}$ must be at least $m$.

Suppose now that $\bar{P}$ is the minimum-weight operator among the even logical operators.
Let $r=O(1)$ be the largest diameter of faces $f\in F$.
Consider two cases: (i) the support of $\bar{P}$ can be partitioned into two disjoint components
separated by a distance greater than $r$; (ii) such a partition does not exist.
In the case (i) we have a decomposition $\bar{P}=\bar{P}_1\bar{P}_2$, where
$\bar{P}_1$, $\bar{P}_2$ individually commute with any face operator
and at least one of $\bar{P}_1$, $\bar{P}_2$ is a non-trivial logical operator.
Note that $\bar{P}_1$, $\bar{P}_2$  cannot be even operators since
it would contradict the weight minimality of $\bar{P}$. Hence both $\bar{P}_1$ and $\bar{P}_2$
are non-trivial odd logical operators. However we have already shown that such operators
must have weight $\Omega(R)$, that is, we arrive at
$|\bar{P}|=|\bar{P}_1|+|\bar{P}_2| =\Omega(R)$. Let us now consider case~(ii).
Since $\bar{P}$ must anti-commute with both $\bar{c}_0$, $\bar{c}_1$, its support must have
odd overlap with both $\gamma_0$ and $\gamma_1$. Condition~(ii) then implies that
$\bar{P}$ must have weight $\Omega(L/r)=\Omega(L)$. In both cases
the diameter of the support of $\bar{P}$ is $\Omega(L)$ since its overlaps with both $\gamma_0$ and $\gamma_1$.

\section{Other constructions of Majorana fermion codes with odd logical operators}
\label{sec:odd}

 In this section we discuss some alternative strategies to construct Majorana fermion codes with odd logical operators.
 Let us begin by considering an unphysical situation when the total number of Majorana fermion modes is {\em odd}.
 Then the operator   $\Call =\prod_{u\in \Lambda} c_u$ is odd and therefore does not belong to $\smaj$.
 On the other hand, $\Call$ has even overlap with any stabilizer and hence $\Call\in \calC(\smaj)\backslash \smaj$,
 that is,  $\Call$ is an odd logical operator.
Note however that when the total number of Majorana fermion modes is odd, we encode
{\em half-integer} number of qubits.  For example, $\Call$ can be the only operator in
${\cal C}(\smaj)\backslash \smaj$. Given the mapping between weakly self-dual CSS codes and
Majorana fermion codes, see Section~\ref{sec:cm}, it is then easy to construct such Majorana fermion codes encoding
half-integer number of qubits. Indeed, let us take any weakly
     self-dual CSS code $[[n,k,d]]$ where the total number of qubits $n$ is odd. Viewed as a Majorana fermion code
     (see Lemma~\ref{fermdoub}) it encodes $k/2$ logical qubits and has $n$ Majorana modes. For example, we could take
      Steane's $[[7,1,3]]$ code~\cite{book:nielsen&chuang} encoding a single qubit.
     The corresponding Majorana fermion code has three generators, $\smaj=\la S_1,S_2,S_3\ra$, where
     $S_1=c_1c_3 c_5 c_7$, $S_2=c_2 c_3 c_6 c_7$, and $S_3=c_4 c_5 c_6 c_7$.
       The logical $\bar{X}$ and $\bar{Z}$ operator
      for the Steane code become a single logical operator
      $\bar{C}_{\rm all}$
      for the Majorana fermion code which encodes half a qubit.
Now we can take another copy of this code, or another weakly self-dual CSS code with an odd number of qubits and take the
product of these codes $\smaj^1 \times \smaj^2$. We now have an even number of Majorana fermion
modes, hence an integer number of encoded qubits. At the same time, odd logical operators
of  $\smaj^1$ and $\smaj^2$ give rise to odd logical operators of the product code
$\smaj^1 \times \smaj^2$.
For two copies of the Steane code, the logical $\bar{X}$ is a weight-$3$ Majorana fermion operator on the $7$ modes of the first
Steane code and the $\bar{Z}$ is the same operator on the $7$ modes of the second Steane code. This plug-and-play procedure
of adding halves of qubits living on separate spatial supports can be enhanced by inserting a piece of passive material which
encodes no qubits between the two coding regions. The (linear) size of this passive region determines the
minimum diameter of even logical operators
$l_{\rm even}$. In this way Kitaev's 1D Majorana fermion model can be viewed as a combination of the trivial code
comprising of a single mode labelled '1', a piece of passive material including  modes $2$ to $2L-1$ in which the
Majorana fermions are  paired, and again a trivial code on mode $2L$.
Such procedure could for example also be applied to another class of 2D color codes, namely the triangular
codes~\cite{BMD:topo} which encode a single qubit and hence half a qubit when the code is viewed as a Majorana fermion code.


\section*{Acknowledgements}
BMT and SB acknowledge support by the DARPA QUEST program under contract number HR0011-09-C-0047. BL acknowledges the financial support and the warm hospitality from IBM Research and its employees.

\section*{Appendix A}

In this Appendix we prove Proposition~\ref{prop:cl}.

\noindent{\bf Proof:}
Given a subset of modes $M\subseteq \Lambda$, we shall define two subgroups of $\smaj$.
The first subgroup denoted as $\smaj(M)$ contains all elements of $\smaj$ whose support is contained in $M$.
The second subgroup denoted as $\smaj^M$ contains all operators $P\in \mjr{2n}$
whose support is contained in $M$ that can be {\em extended} to some stabilizer.
In other words, $P\in \smaj^M$ iff $\supp{(P)}\subseteq M$ and
$PR\in \smaj$ for some operator $R\in \mjr{2n}$ such that $\supp{(R)}\cap M=\emptyset$.
By definition, one has $\smaj(M)\subseteq \smaj^M$.

We shall use the parameterization $\phi\, :\, \mjr{2n} \to \{0,1\}^{2n}$
constructed in Section~\ref{sec:cm}, see Lemma~\ref{fermdoub}.
Consider the linear subspaces (classical codes)
\be
C=\phi(\smaj), \quad C(M)=\phi(\smaj(M)),\quad \mbox{and} \quad
C^M=\phi(\smaj^M).
\ee
By definition, one has the inclusion $C(M) \subseteq C^M$.
Since the code $C$ is weakly self-dual, that is, $C\subseteq C^\perp$, one has
$\sum_{u\in \Lambda} x_u y_u =0$ for all $x\in C(M)$ and $y\in C$.
However, since $x$ has support only on $M$, it translates into $\sum_{u\in M} x_u y_u=0$, that is,
we have also an inclusion
\be
\label{incl}
C(M)\subseteq (C^M)^\perp.
\ee
By abuse of notations, from now on we shall consider the codes $C(M)$ and $C^M$ as linear subspaces
of $\{0,1\}^m$, where $m=|M|$ (note that any vector in $C(M)$ or $C^M$ has all zeros outside of $M$).

There are two possibilities. First, the inclusion Eq.~(\ref{incl}) is an equality, that is,
$C(M)=(C^M)^\perp$. Taking the orthogonal complement of both sides we get
\be
\label{incl=}
C^M=C(M)^\perp.
\ee
Let $\bar{C}\in \calC(\smaj)\backslash \smaj$ be any logical operator and $x=\phi(\bar{C})$.
Decompose $x$ as $x=x_{\rm int}\oplus x_{\rm ext}$, where $x_{\rm int}$ and $x_{\rm ext}$ have support
inside and outside $M$ respectively.
Since $\bar{C}$ commutes with any stabilizer supported on $M$ we conclude that
$x_{\rm int}\in C(M)^\perp$ and hence Eq.~(\ref{incl=}) implies $x_{\rm int}\in C^M$.
It means that $\phi^{-1}(x_{\rm int})$ can be extended to some stabilizer $S\in \smaj$.
Then $\bar{C} S$ acts trivially on $M$. Hence $M$ is cleanable.

The second possibility is that the inclusion Eq.~(\ref{incl}) is strict.
Then there exists some $x\in (C^M)^\perp$ such that $x\notin C(M)$.
Let $\bar{C}=\phi^{-1}(x)$.
Then $\bar{C}$ has support on $M$, commutes with
any element of $\smaj$, but does not belong to $\smaj$.
Hence $\bar{C}$ is a logical operator supported on $M$.

To summarize, we have shown that if $M$ is uncleanable then
the inclusion Eq.~(\ref{incl}) must be strict and hence  there must exist  a logical operator supported on $M$.

Let us now prove the converse. Suppose $\bar{C}$ is a logical operator supported on $M$.
If $\bar{C}$ is odd, then $M$ is uncleanable. Indeed, any stabilizer $S\in \smaj$
must have even overlap with $\supp{(\bar{C})}$ and thus the support of $\bar{C}S$
contains odd number of  modes (and hence at least one)  from $\supp{(\bar{C})}\subseteq M$.
If $\bar{C}$ is even, then there must exist a logical operator $\bar{C}'$ that anti-commutes with $\bar{C}$.
Let us show that $\bar{C}'S$ acts non-trivially on $\supp{(\bar{C})}\subseteq M$ for any stabilizer
$S\in \smaj$ which would imply that $M$ is uncleanable.
Indeed, since $\bar{C}$ is even and anti-commutes with
$\bar{C}'S$, the overlap between $\supp{(\bar{C})}$ and $\supp{(\bar{C}'S)}$ must be odd
and hence $\supp{(\bar{C}'S)}$ contains at least one mode from $\supp{(\bar{C})}\subseteq M$.
Thus in both cases $M$ is uncleanable.
$\Box$

\vspace{2cm}

\bibliographystyle{hunsrt}
\bibliographystyle{apsrev}

\end{document}